\documentclass[twocolumn,aps, prb, showpacs, preprintnumbers, amsmath, amssymb, a4paper, superscriptaddress]{revtex4-1}

\usepackage{graphicx}
\usepackage{dcolumn}
\usepackage{bm}
\usepackage{epstopdf}
\usepackage[utf8]{inputenc}

\usepackage[colorlinks=true,citecolor=black,linkcolor=black,urlcolor=blue,anchorcolor=black]{hyperref}
\usepackage{color}

\usepackage{amsmath}

\begin{document}
\newcommand{\Arg}[1]{\mbox{Arg}\left[#1\right]}
\newcommand{\bb}{\mathbf}
\newcommand{\braopket}[3]{\left \langle #1\right| \hat #2 \left|#3 \right \rangle}
\newcommand{\braket}[2]{\langle #1|#2\rangle}
\newcommand{\be}{\[}
\newcommand{\br}{\vspace{4mm}}
\newcommand{\bra}[1]{\langle #1|}
\newcommand{\braketbraket}[4]{\langle #1|#2\rangle\langle #3|#4\rangle}
\newcommand{\braop}[2]{\langle #1| \hat #2}
\newcommand{\dd}[1]{ \! \! \!  \mbox{d}#1\ }
\newcommand{\DD}[2]{\frac{\! \! \! \mbox d}{\mbox d #1}#2}
\renewcommand{\det}[1]{\mbox{det}\left(#1\right)}
\newcommand{\ee}{\]} 
\newcommand{\eg}{\textbf{\\  Example: \ \ \ }}
\newcommand{\Imag}[1]{\mbox{Im}\left(#1\right)}
\newcommand{\ket}[1]{|#1\rangle}
\newcommand{\ketbra}[2]{|#1\rangle \langle #2|}
\newcommand{\kp}{\arccos(\frac{\omega - \epsilon}{2t})}
\newcommand{\ldos}{\mbox{L.D.O.S.}}
\renewcommand{\log}[1]{\mbox{log}\left(#1\right)}
\newcommand{\Log}{\mbox{log}}
\newcommand{\Modsq}[1]{\left| #1\right|^2}
\newcommand{\nb}{\textbf{Note: \ \ \ }}
\newcommand{\op}[1]{\hat {#1}}
\newcommand{\opket}[2]{\hat #1 | #2 \rangle}
\newcommand{\occ}{\mbox{Occ. Num.}}
\newcommand{\Real}[1]{\mbox{Re}\left(#1\right)}
\newcommand{\so}{\Rightarrow}
\newcommand{\sol}{\textbf{Solution: \ \ \ }}
\newcommand{\thetafn}[1]{\  \! \theta \left(#1\right)}
\newcommand{\tin}{\int_{-\infty}^{+\infty}\! \! \!\!\!\!\!}
\newcommand{\Tr}[1]{\mbox{Tr}\left(#1\right)}
\newcommand{\kb}{k_B}
\newcommand{\rad}{\mbox{ rad}}
\preprint{APS/123-QED}

\title{Valley current generation using biased bilayer graphene dots}

\author{Fionnuala Solomon}
\affiliation{Irish Centre for High-End Computing, 
Grand Canal Quay, Dublin 2, Ireland}
\affiliation{School of Physics, Trinity College Dublin, Dublin 2, Ireland}
\author{Stephen R. Power}
\email{stephen.power@tcd.ie}
\affiliation{School of Physics, Trinity College Dublin, Dublin 2, Ireland}

\date{\today}

\begin{abstract}
Intrinsic and extrinsic valley Hall effects are predicted to emerge in graphene systems with uniform or spatially-varying mass terms.
Extrinsic mechanisms, mediated by the valley-dependent scattering of electrons at the Fermi surface, can be directly linked to quantum transport simulations.
This is a promising route towards more complete experimental investigation of valleytronic phenomena in graphene, but a major obstacle is the difficulty in applying the sublattice-dependent potentials required.
Here we show that strongly valley-dependent scattering also emerges from bilayer graphene quantum dots, where the gap size can be easily modulated using the interlayer potentials in dual-gated devices.
Robust valley-dependent scattering and concomitant valley currents are observed for a range of systems, and we investigate the role of dot size, mass strength and additional potential terms.
Finally, we note that a strong valley splitting of electronic current also emerges when a biased bilayer dot is embedded in a single layer of graphene, but that the effect is less robust than for a bilayer host. 
Our findings suggest that bilayer graphene devices with custom mass profiles provide an excellent platform for future valleytronic exploration of two-dimensional materials.
\end{abstract}

\maketitle

\section{Introduction}
Valleytronics\cite{schaibley2016valleytronics} is an emerging field in which the relative occupation of inequivalent local extrema, or \emph{valleys}, in the band dispersion of a material can be exploited to encode, transport and process information in a similar manner to the \emph{spin} degree of freedom in spintronics.
Aside from the fundamental interest in harnessing a new quantum degree of freedom, valleytronic components could also play key roles in future quantum computing technologies \cite{rohling2012universal, PhysRevLett.108.126804, laird2013valley}.
Graphene, alongside other two-dimensional (2D) materials, is a promising candidate in this regard due to the presence of valleys formed at the Dirac points, $K$ and $K$'.
However, a key obstacle is the absence of valleytronic analogues to magnetic fields and ferromagnetic contacts, with which to manipulate and detect valley-polarized currents.\cite{Cresti2016}
In many cases, electrons from each valley behave identically and contribute equally, so that resolving valley-related behaviour using electronic measurements is not possible.
In materials with broken inversion symmetry, electrons from individual valleys can be excited by circularly polarised light of different chiralities, allowing optoelectronic access to the valley degree of freedom. \cite{xiao2012coupledSV, cao2012valley, li2014valleysplitting,PhysRevB.99.115441}
However, for device applications, an all-electronic control of valley properties is highly desireable. \cite{PhysRevB.96.245410}
Bespoke defects or strains have been theoretically proposed to allow useful functionalities such as valley filtering, but the corresponding experimental implementation remains challenging. \cite{rycerz2007valley, garcia-pomar2008valley, fujita2010valley, gunlycke2011valley, chen2014valley, PhysRevB.96.201407, PhysRevApplied.11.044033, levy2010strain, PhysRevB.77.205421, guinea2010energy, vozmediano2010gauge,settnes2016pmandtriaxial, qi2013resonant,settnes2016graphenebub,milovanovic2016strain,settnes2017valleygauge,Zhai2018,Stegmann_2018,wu2018foldwgs, PhysRevB.99.035411, JPSJ.88.083701, PhysRevApplied.11.054019, 2019arXiv190804604T, PhysRevB.103.115437}

Promising signatures of valley phenomena have instead emerged from non-local resistance ($R_\mathrm{NL}$) measurements in commensurately stacked graphene/hexagonal boron nitride (hBN) systems \cite{gorbachev2014detecting}.
Large $R_\mathrm{NL}$ signals here have been interpreted in terms of an intrinsic valley Hall effect (VHE), driven by a bulk Berry curvature induced by a \emph{mass} term (\emph{i.e.} sublattice asymmetry) arising from the interaction between the graphene and hBN layers. \cite{PhysRevLett.114.256601, xiao2007valley, ando2015theory, PhysRevB.94.121408}
Under this argument, a valley Hall conductivity within the band gap generates a long-ranged valley current, enhancing $R_\mathrm{NL}$ beyond standard ohmic contributions. \cite{gorbachev2014detecting,Song2015PNAS}.
However, this interpretation of experimental $R_\mathrm{NL}$ has been questioned by quantum transport simulations \cite{kirczenow2015valleyNL, Cresti2016, MarmolejoTejada2018} and subsequent experiments \cite{zhu2017edge, aharonsteinberg2020longrange}.
In particular, this mechanism relies on contributions from electrons in the Fermi sea, whereas typically device measurements are dictated only by electrons at the Fermi surface, which will be exponentially suppressed in a gapped system. \cite{kirczenow2015valleyNL, MarmolejoTejada2018}
The relative importance of edge or bulk, and topological or non-topological, contributions is still strongly debated. \cite{zhu2017edge, MarmolejoTejada2018, komatsu2018observation, Brown2018, Song2019, aharonsteinberg2020longrange, li2020topological} 
These issues can be circumvented if we shift our focus from global to local mass terms.
Embedding `mass dots' into an otherwise pristine graphene sheet induces an extrinsic valley Hall effect that emerges from valley-dependent scattering at each dot,\cite{Aktor2021} similar to skew scattering mechanisms for spin and valley Hall effects. \cite{aires2014skew, aires:extrinsicSHE, PhysRevB.91.165407, PhysRevB.96.201407}.
This effect is mediated by Fermi surface electrons, and quantum transport simulations reveal an enhancement of $R_\mathrm{NL}$ which can be directly connected to the flow of a valley-polarized current throughout the device.
While the Moir\'{e} pattern in commensurate graphene/hBN structures naturally introduces a non-uniform mass profile\cite{woods2014commensurate, jung2015origin}, the combination of spatially-varying mass, potential and strain fields, together with the possibility of non-topological edge currents,\cite{MarmolejoTejada2018} will make it difficult to attribute experimental signatures to a single mechanism.

To investigate scattering-induced valley splitting in the absence of additional effects, we ideally require a tuneable mass inside the dot and a pristine system outside.
This is difficult to achieve in single layer graphene (SLG) using either substrate effects as discussed above or other methods, such as sublattice-asymmetric doping. \cite{zhao2011visualizing, lv2012nitrogen, usachov2016large, zabet2014segregation, lawlor2014sublattice, lawlor2014sublattice2, lherbier2013electronic, aktor2016electronic} 
However, the possibility of creating exactly this type of selectively-gapped structure emerges if we move to bilayer graphene (BLG) and consider an asymmetry in layer instead of sublattice potentials.
It is long established that applying an interlayer potential to Bernal-stacked BLG opens a band gap in its low energy band structure. \cite{McCann2006, Castro2007, McCann2013}
There are also promising indications that this kind of biased BLG system can display interesting valleytronic behaviour.
An analogous non-local resistance behaviour to that discussed for SLG above is also reported for globally-gapped BLG devices and interpreted using valley-dependent Berry curvature mechanisms.\cite{sui2015gate, shimazaki2015generation, endo2019topological}  
Previous theoretical works have considered scattering from various bilayer barrier geometries and predict that the electronic transmission from either SLG\cite{linearBL} or BLG\cite{Schomerus_2010} into a biased bilayer region displays a valley dependence.

Given the parallels with a mass term in SLG, in this work we consider the valley-dependent scattering from locally-biased dots in a bilayer sheet.
We find that they provide a robust platform in which to induce tunable valley-splitting, valley currents and an extrinsic valley Hall effect.
In Section \ref{secmethods}, we outline the theoretical methods used to solve the scattering problem using the 4-band continuous Dirac model.
The energy, valley and angular dependence of scattering from a typical biased BLG dot are outlined in Section \ref{secresults_scat}.
We find an incoming unpolarized electron wave experiences strong valley-dependent scattering over a wide energy range within the band gap of the biased region. 
The roles of the dot size, mass strength and an additional shift of the band centre in the biased dot are the discussed in Sections \ref{sec_size_strength} and \ref{sec_asymm}.
We also consider the possibility of inducing valley currents by biasing bilayer islands within a sheet of SLG. 
This is discussed in Sec. \ref{sec_SLG}, where we find promising behaviour in certain cases but note a strong dependence on the edge geometry of the bilayer regions.

Our findings show that biased bilayer dots are an extremely promising route towards enabling valleytronic functionality in graphene.
They induce a robust valley splitting which can be easily controlled using electric fields.
Furthermore, unlike other strain-based proposals, this setup does not require a deformation of the system.
In contrast to substrate-induced mass terms, the magnitude of the gap can be tuned to study the effect over a wider parameter range.
This approach offers a clear advantage over globally-gapped systems which rely on non-local measurements, which are difficult to reconcile with quantum transport simulations, to infer valleytronic behaviours.
Valley-dependent scattering from biased dots will give rise to valley-polarized currents at the Fermi surface, which can be directly compared to device simulations.
The transport signatures of devices containing either individual dots or superlattices will give distinct and direct fingerprints of the predicted valleytronic behaviour.
Finally, we note that the dual-gate experimental setup required to create finite regions of biased BLG have recently been used to create quantum dots\cite{dotsKurzmann2019, dotsKurzmann2019a, dotsBanszerus2020, dotsBanszerus2020a, dotsGe2020, dotsGarreis2021} and quantum point contacts\cite{Overweg2018, Kraft2018, Lee2020}.
Interestingly, the application of an external magnetic field in the latter case leads to a valley splitting of the energy levels in the system, but the mechanism involved is different to that in the current work.

\section{Method} \label{secmethods}
\begin{figure}
\includegraphics[width=0.48\textwidth]{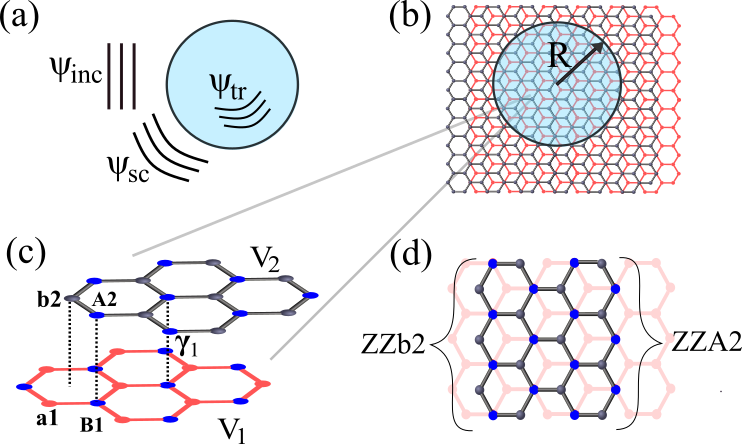}
	\caption{(a) Setup of scattering problem, showing the incoming plane wave and the scattered wave outside the dot, and the transmitted wave in the dot.  (b),(c) Bilayer geometry, including the dot region where the layer potentials $V_1$ and $V_2$ are applied, and the individual sites in the unit cell. (d) Two possible edge geometries if the top layer is finite.}
	\label{setup}
\end{figure}

In an AB-stacked bilayer,  the layers have a small in-plane shift relative to each other such that that atoms from the $A$ sublattice on the top layer lie directly above the $B$ atoms from the bottom layer, as shown in Fig. \ref{setup}(b).
These two sites, which we denote A2 and B1 respectively in Fig. \ref {setup}(c), are collectively referred to as \emph{dimer} sites.
The remaining b2 and a1 sites lie directly opposite the hexagon centres and are referred to as \emph{non-dimer} sites.
In this notation, an uppercase (lowercase) sublattice index indicates whether a site is a dimer (non-dimer) and the numerical index refers to the bottom (1) or top (2) layer.
The electronic structure of biased AB-stacked bilayer graphene is given by the Dirac Hamiltonian
\begin{equation}
    \mathcal{H}(k) = \hbar v_F \begin{pmatrix}
    \tilde{V_2} & \tau k e^{-i\tau \theta} & 0 & \tilde{\gamma_1}\\
    \tau k e^{i\tau\theta} & \tilde{V_2} & 0 & 0 \\
    0 & 0 &\tilde{V_1}&\tau k e^{-i\tau\theta} \\
    \tilde{\gamma_1} & 0 & \tau k e^{i\tau\theta} & \tilde{V_1}
    \end{pmatrix}
    \label{hamiltonian}
\end{equation}
in the vicinity of the $K$ and $K^\prime$ points. 
This matrix, and the associated 4-element spinors discussed below, are ordered (A2, b2, a1, B1).
Here, $V_1$ and $V_2$ represent the onsite potentials on the two layers (Fig. \ref{setup}(c)), which can be separately controlled in a dual-gated setup, $\gamma_1$ is the direct intralayer hopping between AB dimer sites and $\tau = \pm 1$ is the valley index. 
We work with scaled variables $\tilde{X} = \frac{X}{\hbar v_F}$ to simplify the algebra.
Higher-order skew hopping terms between dimer and non-dimer sites in different layers are omitted, as they are not expected to play a significant role in the low-energy band structure.
However, such terms are necessary to include higher-order effects, such as trigonal warping and minivalley formation.\cite{McCann2013, PhysRevB.98.155435}
The wavevector $\mathbf{k} = (k_x, k_y) = (k \cos{\theta}, k \sin{\theta})$ is related to the energy $E$ by the dispersion relation 
\begin{multline}
    \tilde{E}(k) = \frac{\tilde{\delta}}{2} + \eta_1   \left[ \frac{\tilde{\gamma_1}^2}{2} + \left(\frac{\tilde{\Delta}}{2}\right)^2 +  k^2  \right.
    \\  \left. + \eta_2 \sqrt{ \frac{\tilde{\gamma_1}^4}{4} +  k^2 \left( \tilde{\gamma_1}^2 + \tilde{\Delta}^2 \right)  } \, \right]^{\frac{1}{2}} \,.
    \label{eq:V1V2dispersion}
\end{multline}

Eq. \eqref{eq:V1V2dispersion} gives rise to two valence ($\eta_1=-1$) and two conduction ($\eta_1=+1$) bands, with the band centre coinciding with the half-filling Fermi energy and given by the the sum of layer potentials $\delta = V_1 + V_2$.
In the absence of an interlayer potential, $\Delta=V_1-V_2 =0$, a pair of low energy bands ($\eta_2 = -1$) touch at the Dirac point, where they are entirely localised on non-dimer sites.
A non-zero $\Delta$ acts like a mass term and opens a gap in these low energy bands.
In addition to opening a gap, large values of $\Delta$ modify the shape of these bands into a characteristic `Mexican Hat' shape, where the minimum gap $\tfrac{|\Delta| \gamma_1}{\sqrt{ \Delta^2 + \gamma_1}}$ occurs a small distance from the Dirac points.
A pair of higher energy bands ($\eta_2 = +1$) are separated by a pseudogap of width $\approx\gamma_1$ at the Dirac point.
The corresponding wavevectors for a given energy $E$ are given by
\begin{equation}
    k^\pm  = \sqrt{ \tilde{\epsilon}_c^2 + \left(\frac{\tilde{\Delta}}{2}\right)^2 \mp \tilde{\gamma}_1 \sqrt{ \tilde{\epsilon}_c^2 \left(1 + \frac{\tilde{\Delta}^2}{\tilde{\gamma}_1^2}\right) - \left(\frac{\tilde{\Delta}}{2}\right)^2 } } \label{eqkpm} 
\end{equation}
where $\epsilon_c = E - \frac{\delta}{2}$ is the energy measured from the bandcentre, and $k^-$ and $k^+$ correspond respectively to the lower ($\eta_2=-1$) and higher ($\eta_2=+1$) energy bands.
The associated eigenfunctions can be written\cite{bl_step}
\begin{equation}
    \psi (\mathbf{k}, \mathbf{r})    = A  
    \begin{pmatrix} \tilde{\gamma_1}  (\tilde{E} - \tilde{V_1}) (\tilde{E} - \tilde{V_2})\\
    \tau \tilde{\gamma_1} (\tilde{E} - \tilde{V_1})  k e^{i\tau\theta}\\
    \tau \left[ (\tilde{E} - \tilde{V_2})^2 - k^2 \right] k e^{-i\tau\theta}\\ 
    \left[ (\tilde{E} - \tilde{V_2})^2 - k^2 \right] (\tilde{E} - \tilde{V_1}) \end{pmatrix}
    e^{i\mathbf{k}.\mathbf{r}} \,.
    \label{eq:psiV1V2}
\end{equation}
up to a normalisation constant $A$. 

To investigate how biased BLG nanostructures can give rise to valley-dependent currents, we consider the scattering of an incoming electron plane wave from a circular biased dot, as shown in Fig. \ref{setup}(a),(b). 
Outside the dot, we set $V_1=V_2=0$, which results in simplified expressions for the dispersion relation, wavevectors and eigenfunctions
\begin{align}
     \tilde{E}_0(k) & = \eta_1 \left[ \sqrt{\left(\frac{\tilde{\gamma}_1}{2}\right)^2 +k^2} + \eta_2 \frac{\tilde{\gamma}_1}{2} \right] \label{eqE0}\\
     k^\pm_0 & = \sqrt{|\tilde{E}| ( |\tilde{E}| \mp \tilde{\gamma_1} ) }  \label{eq:k0s}\\
     \psi_0 (\mathbf{k}, \mathbf{r})  & = \frac{1}{\sqrt{2 \, (\tilde{E}^2 + |k|^2)}}  
    \begin{pmatrix} |\tilde{E}|\\
    \eta_1 \tau k e^{i\tau\theta}\\
    \eta_2 \tau k e^{-i\tau\theta}\\ 
    \eta_1 \eta_2 |\tilde{E}| \end{pmatrix}
    e^{i\mathbf{k}.\mathbf{r}} \,.
\end{align}
In this region,  $k_0^-$ is real for all values of energy, whereas $k_0^+$ is purely imaginary within the gap of the higher energy bands, \emph{e.g.} $-\gamma_1 < \epsilon_c < \gamma_1$.
Thus, depending on the energy, we have either two propagating contributions, or one propagating and one evanescent contribution, to the total wavefunction.
Inside the dual-gated dot, both contributions are evanescent if $|\epsilon_c| < |\frac{\Delta}{2}|$, \emph{i.e.} if the energy lies in the gap induced by the gates.
We are primarily interested in the experimentally achievable regime $|E| \lesssim |\Delta| \ll \gamma_1$, where low energy electrons are scattered by gate-induced potentials considerably smaller in magnitude than the interlayer coupling. 
The scattering problem is solved by wavefunction matching at the dot interface.
The wavefunction outside the dot consists of both incident ($\psi_\mathrm{inc}$) and scattered ($\psi_\mathrm{sc}$) terms, which in principle each contain two contributions, corresponding to the two possible wavevectors $k_0^\pm$.
However, we take the incident wave to be a plane wave approach from $x= -\infty$, so in the relevant energy window the term associated with $k_0^+$ is evanescent and can be ignored as it has decayed exponentially before reaching the vicinity of the dot.
The scattered wave, on the other hand, can contain both propagating and evanescent contributions, so both $k_0^\pm$ terms must be included when matching the wavefunction the interface.
Similarly, the transmitted wave $\psi_\mathrm{tr}$ inside the dot contains contributions from both $k^\pm$ terms, one or both of which are evanescent depending on the layer potentials.
Therefore, at the boundary of a dot of radius $R$, the wavefunction matching condition reads
\begin{multline}
\psi_\mathrm{inc} (k_0^-, r=R) + \psi_\mathrm{sc} (k_0^-,  r=R)  + \psi_\mathrm{sc} (k_0^+,  r=R) \\ = \psi_\mathrm{tr} (k^-,  r=R) + \psi_\mathrm{tr} (k^+,  r=R) 
\label{eq:wavefnmatch}
\end{multline}

For a circular dot, it is convenient to rewrite the problem in terms of polar coordinate operators.
Using $\mathbf{k} = -i \vec\nabla$ and
\begin{align}
    \frac{\partial}{\partial x} & = \cos \theta \frac{\partial}{\partial r} - \frac{1}{r} \sin \theta \frac{\partial}{\partial \theta} \\
    \frac{\partial}{\partial y} & = \sin \theta \frac{\partial}{\partial r} + \frac{1}{r} \cos \theta \frac{\partial}{\partial \theta}
\end{align}
we can rewrite the Hamiltonian as
\begin{widetext}
\begin{equation}
    \mathcal{H}(k) = \hbar v_F \begin{pmatrix}
     \tilde{V_2} &  e^{-i\tau\theta}  \left(-i \tau \partial_r -\frac{1}{r} \partial_\theta\right)  & 0 & \tilde{\gamma_1}\\
    e^{i\tau\theta}  \left(-i \tau \partial_r +\frac{1}{r} \partial_\theta\right) &  \tilde{V_2} & 0 & 0 \\
    0 & 0 &  \tilde{V_1} & e^{-i\tau\theta}  \left(-i \tau \partial_r -\frac{1}{r} \partial_\theta\right) \\
    \tilde{\gamma_1} & 0 & e^{i\tau\theta}  \left(-i \tau \partial_r +\frac{1}{r} \partial_\theta\right) &  \tilde{V_1}
    \end{pmatrix}
    \label{hamiltonian_radial}
\end{equation}
\end{widetext}

Since this Hamiltonian commutes with the angular momentum operator, a standard approach is to expand wavefunctions in terms of angular momentum basis states, which are typically expressed in terms of Bessel functions\cite{Nilsson2008, airesf2011unified, heinisch2013, schulz2015electronflow, Peterfalvi2009, Aktor2021}. 
A plane wave incident along the $x$-direction can be written in this way by exploiting the identity
\begin{equation}
e^{i k x} = e^{i k r \cos \theta} =\sum_{m=-\infty}^{\infty} i^m J_m (k r) \, e ^ {i m \theta}
\end{equation}
For example, to use this approach for the 4-element bilayer spinor wavefunction in the region outside the dot, we consider an Ansatz solution of the form 
\begin{equation}
    \psi_{m}    =  
    \begin{pmatrix} | \tilde{E} | \, A (r) \, e^{i m \theta} \\
    i \eta_1 k \, B (r) \, e^{i (m+\tau) \theta} \\
    - i \eta_2 k \, C (r) \, e^{i (m-\tau) \theta}\\ 
    \eta_1 \eta_2 | \tilde{E} | \, D (r) \, e^{i m \theta} \end{pmatrix}
    \,,
    \label{ansatz}
\end{equation}
and substitute this into the Schroedinger equation using Eq. \eqref{hamiltonian_radial} with  $\tilde{V_1} =  \tilde{V_2} = 0$ to give the system of equations
\begin{align}
    \left[ \tau \partial_r + \frac{(m+\tau)}{r}\right] k B (r) + \eta_2 \tilde{\gamma_1} | \tilde{E} | D (r) & =  \tilde{E}^2 \, A (r) \label{sys1} \\
    \left[ -\tau \partial_r + \frac{m}{r}\right] A (r) & =  k B (r) \label{sys2} \\
    \left[ \tau \partial_r + \frac{m}{r}\right] D (r) & =  k C (r) \label{sys3} \\
    \left[ - \tau \partial_r + \frac{(m-\tau)}{r}\right]  k C (r) +  \eta_2 \tilde{\gamma_1} | \tilde{E} | A (r) & = \tilde{E}^2 D (r) \label{sys4}
\end{align}
Further substituting Eqs. \eqref{sys2} and \eqref{sys3} into Eqs. \eqref{sys1} and \eqref{sys4} respectively yields
\begin{align}
    \left[ -\partial_r^2  -\frac{1}{r} \partial_r + \frac{m^2}{r^2} - \tilde{E}^2 \right] A (r)  & = - \eta_2 \, \gamma_1 \, | \tilde{E} | \, D (r) \label{sys5} \\
    \left[ -\partial_r^2  -\frac{1}{r} \partial_r + \frac{m^2}{r^2} - \tilde{E}^2 \right] D (r)  & = - \eta_2 \, \gamma_1 \, | \tilde{E} | \, A (r) \label{sys6}
\end{align}
with a final substitution of Eq. \eqref{sys5} into Eq. \eqref{sys6} giving an equation for $A(r)$ only
\begin{equation}
    \left[ -\partial_r^2  -\frac{1}{r} \partial_r + \frac{m^2}{r^2} - \tilde{E}^2 \right] ^2  A (r) = \left(\gamma_1 \,  \tilde{E} \right)^2 A (r)
\end{equation}
which can be rewritten as 
\begin{equation}
    \left[ \rho^2 \partial_\rho^2  + \rho \partial_\rho + \left( \rho^2 - m^2\right) \right] A (\rho) = 0
    \label{Bessel}
\end{equation}
where $\rho = k r$ and $k = k_0^\pm(E)$, \emph{i.e.} the wavevector solutions from Eq. \eqref{eq:k0s}.
Eq. \eqref{Bessel} is the Bessel equation, whose solutions are Bessel functions $J_m(\rho)$.
$D(r)$ can be expressed similarly, whereas $B(\rho)$ and $C(\rho)$ are associated with Bessel's functions of order ${m+\tau}$ and $m-\tau$ respectively.
Therefore wavevectors of the type 
\begin{equation}
    \psi_m^{0,J} (kr, \theta)     =  
    \begin{pmatrix} | \tilde{E} | \, J_m (k r) \, e^{i m \theta} \\
    i \eta_1 k \, J_{m+\tau}(k r) \, e^{i (m+\tau) \theta} \\
    - i \eta_2 k \,J_{m-\tau}(k r) \, e^{i (m-\tau) \theta}\\ 
    \eta_1 \eta_2 | \tilde{E} | \, J_m (k r) \,  e^{i m \theta} \end{pmatrix}
    \,,
    \label{eq:psimJdef}
\end{equation}
are solutions to the Hamiltonian in Eq. \eqref{hamiltonian_radial}.
Note that different sign choices in the Ansatz in Eq. \eqref{ansatz} can give rise to the same system of Bessel equations, but not all of these will solve the required Schrodinger equation. 
To confirm that Eq. \eqref{eq:psimJdef} is a valid solution, it is substituted into the Schrodinger equation, and the resulting system of equations are checked using the Bessel function identities
\begin{equation}
\begin{aligned}
   \frac{\partial J_{m}(z)}{\partial z } & = J_{m-1}(z) - \frac{m}{z} J_m(z) \\
   & = \frac{m}{z} J_m(z) - J_{m+1}(z) \,.
   \end{aligned}
\end{equation}

An incident plane wave, propagating along the positive $x$ direction, can be expanded as
\begin{align}
    \psi_\mathrm{inc} (k, \mathbf{r})  & = \frac{1}{\sqrt{2 \, (\tilde{E}^2 + |k|^2)}}  
    \begin{pmatrix} |\tilde{E}| \\
    \eta_1 \tau k  \\
    \eta_2 \tau k  \\ 
    \eta_1 \eta_2 |\tilde{E}| \end{pmatrix}
    e^{i k x} \\
    & = \frac{1}{\sqrt{2 \, (\tilde{E}^2 + |k|^2)}} \sum_{m=-\infty}^{\infty} i^m \psi_m^{0,J} (kr, \theta) \,. 
\end{align}
The functions $ \psi_m^{0,J} (kr, \theta)$ are composed of a linear combination of eigenvectors of the total angular momentum operator, including orbital and pseudospin contributions.
Since this operator commutes with the Hamiltonian, the functions $\psi_m^{0,J} (kr, \theta)$ provide a good basis in which to expand any wave.
Both components of the scattered wave ($k=k_o^\pm$) can be expanded 
\begin{equation}
    \psi_\mathrm{sc} (k, \mathbf{r})  = \frac{1}{\sqrt{2 \, (\tilde{E}^2 + |k|^2)}} \sum_{m=-\infty}^{\infty} c^\mathrm{sc}_{m,k} \,  i^m \, \psi_m^{0,H} (kr, \theta) \,,
\end{equation}
where $c^\mathrm{sc}_{m,k}$ is the scattering coefficient and $\psi_m^{0,H} (kr, \theta)$ is defined as in Eq. \eqref{eq:psimJdef}, but with the Bessel functions replaced by Hankel functions $H_m (kr)$, which have the required  asymptotic behaviour $H_m (kr \rightarrow \infty) \rightarrow 0$.

The transmitted wavefunction inside the dot can be similarly expanded, but considering the full Hamiltonian in Eq. \eqref{hamiltonian} including non-zero $V_1$ and $V_2$,  which yields
\begin{equation}
    \psi_\mathrm{tr} (k, \mathbf{r})  = \sum_{m=-\infty}^{\infty} c^\mathrm{tr}_{m,k} \,  i^m \, \psi_m^{V,J} (kr, \theta) \,,
    \label{transmitted}
\end{equation}
where $c^\mathrm{tr}_{m,k}$ are the transmission coefficients for mode $m$ for the two possible wavevectors $k=k^\pm$ given by Eq. \eqref{eqkpm} .
The form $\psi_m^{V,J} (kr, \theta)$ is found following a similar procedure to that outlined for $\psi_m^{0,J}$ in Eqs. \eqref{ansatz} -- \eqref{eq:psimJdef}  
\begin{equation}
    \psi_m^{V,J} (kr, \theta)     =  
    \begin{pmatrix} \tilde{\gamma_1}  (\tilde{E} - \tilde{V_1}) (\tilde{E} - \tilde{V_2}) \, J_m (k r)  e^{i m \theta} \\
    i \tilde{\gamma_1} (\tilde{E} - \tilde{V_1})  k \, J_{m+\tau}(k r)  e^{i (m+\tau) \theta} \\
    - i \left[ (\tilde{E} - \tilde{V_2})^2 - k^2 \right] k  \,J_{m-\tau}(k r)  e^{i (m-\tau) \theta}\\ 
    \left[ (\tilde{E} - \tilde{V_2})^2 - k^2 \right] (\tilde{E} - \tilde{V_1}) \, J_m (k r)   e^{i m \theta} \end{pmatrix}
    \,.
    \label{eq:psimJdefV}
\end{equation}
We note that we do not explicitly include a normalisation constant, analogous to the $\frac{1}{\sqrt{2 \, (\tilde{E}^2 + |k|^2)}}$ term in the incoming and scattered waves.
This is absorbed into the $c^\mathrm{tr}_{m,k}$ coefficients when the total wavefunctions at each side of the dot boundary are matched.

\begin{figure}
\includegraphics[width=0.48\textwidth]{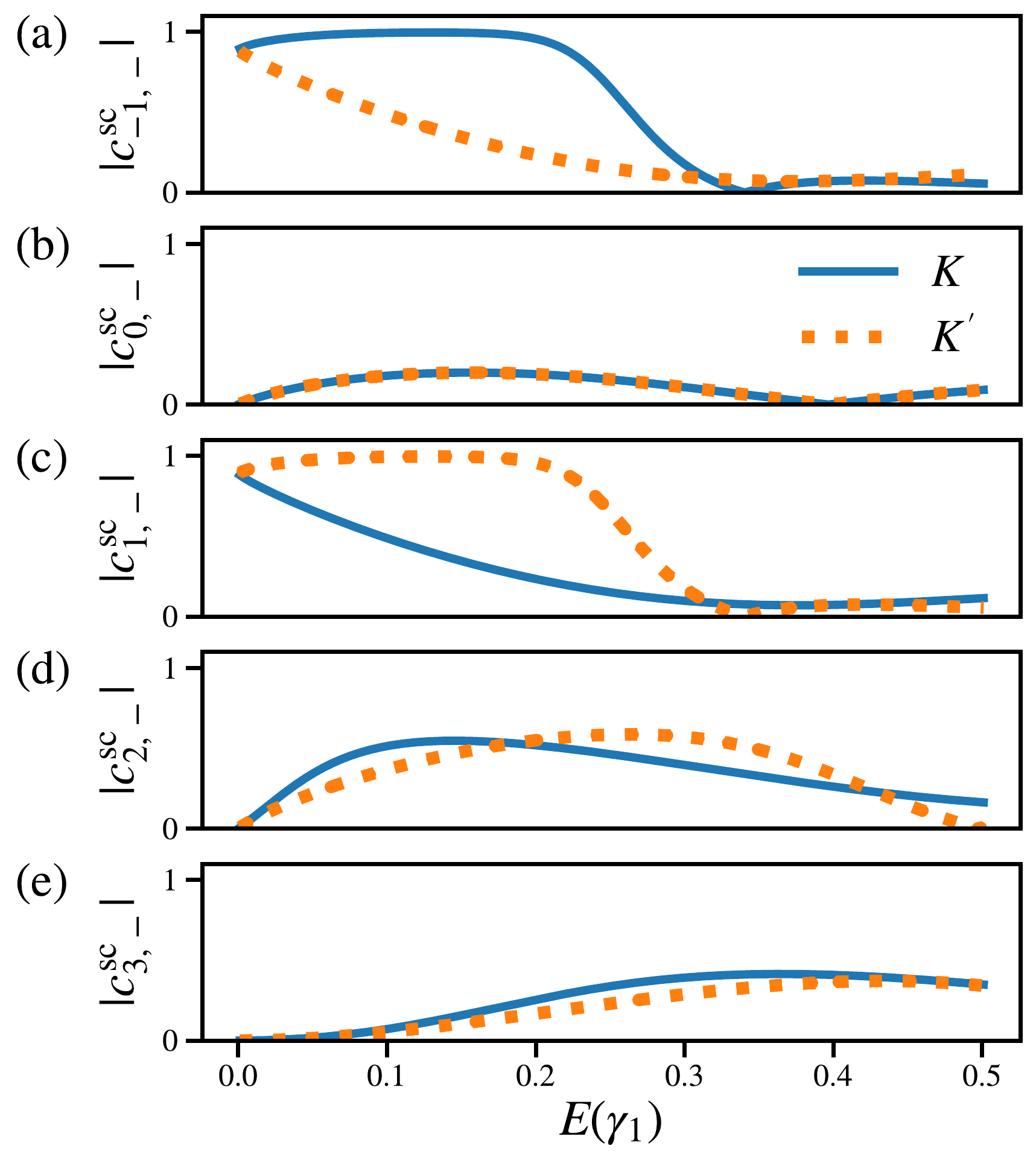}
	\caption{(a)-(e) Magnitude of the scattering coefficients $c^\mathrm{sc}_{m,-}$ for both valleys, for $m=-1, \dots, 3$, for a dual-gated dot with $R=5.0 l_0$ and $V_1=-V_2=0.2\gamma_1$. 
	At low energies, only a small number of modes near $m=0$ contribute to the total wavefunction. 
	Note that  $c^\mathrm{sc}_{m,-} (K) = c^\mathrm{sc}_{-m,-} (K^\prime)$, so that each $K$ valley mode has a corresponding $K^\prime$ mode with equal magnitude and opposite angular momentum.}
	\label{figmodes}
\end{figure}

The complete scattering problem for a circular dot can then be solved by writing an equation like Eq. \ref{eq:wavefnmatch} for each mode $m$ and calculating the full set of scattering and transmitted coefficients $c^\mathrm{sc}_{m,-}$, $c^\mathrm{sc}_{m,+}$, $c^\mathrm{tr}_{m,-}$, $c^\mathrm{tr}_{m,+}$. 
In principle the sum over $m$ required to calculate the full wavefunction in each region runs from $-\infty$ to $\infty$, but as we discuss below, this can be restricted to a finite number of modes around $m=0$.
The problem is solved explicitly for each valley, but the solutions are closely related. 
Fig. \ref{figmodes} shows, for a sample dot,  the magnitude of the scattering coefficients $|c^\mathrm{sc}_{m,-}|$ for $m=-1, \dots, 3$, with the $K$ ($K^\prime$) version of the mode represented by a solid (dotted) line.
In general, only a small number of modes near $m=0$ contribute at low energies, with higher order modes entering gradually as the energy increases.
There is a close connection between the $m$ and $-m$ modes, which can be seen by comparing Fig. \ref{figmodes}(a) and (c).
The coefficients obey $c^\mathrm{sc}_{m,-} (K) = c^\mathrm{sc}_{-m,-} (K^\prime)$, so that each $K$ mode has a corresponding $K^\prime$ mode which contributes with equal magnitude and opposite angular momentum. \cite{Aktor2021}

From the full wavefunction for a particular valley,
\begin{equation}
    \Psi^\tau  = 
    \left\{\begin{array}{lr}
        \psi_\mathrm{tr}^\tau & \text{for } r\le R\\
        \psi_\mathrm{inc}^\tau + \psi_\mathrm{sc}^\tau & \text{for } r> R
        \end{array} \right. \,,
\end{equation}
we can calculate the electron density $n^\tau$ and probability current $j^\tau$ associated with that valley throughout the system using
\begin{align}
    n^\tau & = \Psi^{\tau\dagger}  \Psi^{\tau} \label{eq:occwf}\\
    j^\tau & = v_f \, \Psi^{\tau\dagger} \, (I_2 \otimes \sigma_{\tau}) \, \Psi^{\tau} \,, \label{eq:currwf}
\end{align}
where $\sigma_{\tau} = (\tau\sigma_x, \sigma_y)$ and  $I_2$ is the identity matrix in layer space.
The total density or current is calculated by summing the two valley contributions, whereas the valley quantity is given by the difference between the $K$ and $K^\prime$ contributions.

In discussing the overall scattering or valley-splitting characteristics of a dot, it is useful to consider just the scattered current, and to examine the radial component which can be calculated by replacing $\sigma_\tau$ in Eq. \eqref{eq:currwf} with  
\begin{equation}
    \sigma_{\tau}^{\text{rad}}= \tau\sigma_1\cos{\theta}+\sigma_2\sin{\theta} \,
\end{equation}
In the far-field limit, we can make use of the asymptotic behaviour of Hankel functions
\begin{equation}
    \lim_{r\to\infty} H_m(kr)=\sqrt{\frac{2}{\pi kr}}e^{ikr}i^{-(m+1)} \,.
\end{equation}
Furthermore, in the most relevant energy range $|E| < \gamma_1$, we need only consider scattered current contributions from the $k_0^-$ mode in the region far from the dot, as the contributions of the $k_0^+$ are exponentially suppressed.
We find
\begin{equation}
    j^{\tau, {\text{rad}} }_{r\to \infty} = \frac{4 v_F \eta_1 |E|}{\pi r (E^2 + |k_0^-|^2)} \sum_{m, n=-\infty}^{\infty}  c^\mathrm{sc, *}_{m,-} c^\mathrm{sc}_{n,-} e^{i (n-m) \theta}
\end{equation}
The scattering efficiency $Q$ is related to the scattering cross section $\sigma$ and can be calculated from the far-field scattered current by considering all possible scattering angles
\begin{equation}
\begin{aligned}
    Q   = \frac{\sigma}{2R} & =  \frac{1}{2R} \int_0^{2\pi} r \, j^{\tau, {\text{rad}} }_{r\to \infty} (r, \theta) \; \mathrm{d} \theta \\
     & = \frac{4 v_F \eta_1 |E|}{(E^2 + |k_0^-|^2) R} \sum_{m=-\infty}^{\infty}  |c^\mathrm{sc}_{m,-}|^2
     \label{eqQ}
\end{aligned}
\end{equation}
As we will observe explicitly below, the scattering of $K$ and $K^\prime$ electrons by dual-gated dots is anti-symmetric around the $x$-axis.
This emerges from the equal magnitude counter-propagating flows generated by scattered modes of equal strength and opposite angular direction, as discussed above.
Therefore a preferential deflection of $K$ electrons to the upper-half plane is always compensated by an equal deflection of $K^\prime$ electrons to the lower half plane.
The scattering efficiency $Q$ is therefore identical for both valleys.
To measure the valley-scattering efficiency of a dot, we instead consider the difference between the $K$ and $K^\prime$ contributions to the far-field scattered current over the upper half-plane, and define
\begin{equation}
\begin{aligned}
    \xi^v & = \frac{1}{2R} \,  \int_{0} ^{\pi}  r \left( j^{K, {\text{rad}} }_{r\to \infty} (\theta)  -  j^{K^\prime, {\text{rad}} }_{r\to \infty} (\theta) \right)\, \mathrm{d} \theta \\
    & = \frac{8 v_F \eta_1 |E| i}{\pi (E^2 + |k_0^-|^2) R} \sum_{m, n=-\infty}^{\infty} \frac{c^\mathrm{sc, *}_{m,-} c^\mathrm{sc}_{n,-} \sin^2 \left( \frac{(n-m)\pi}{2} \right) }{n-m} \,.
    \label{eqxi}
\end{aligned}
\end{equation}

\section{Results} \label{secresults}

\subsection{Scattering from dual-gated dots in bilayer graphene}
\label{secresults_scat}

\begin{figure}
\includegraphics[width=0.48\textwidth]{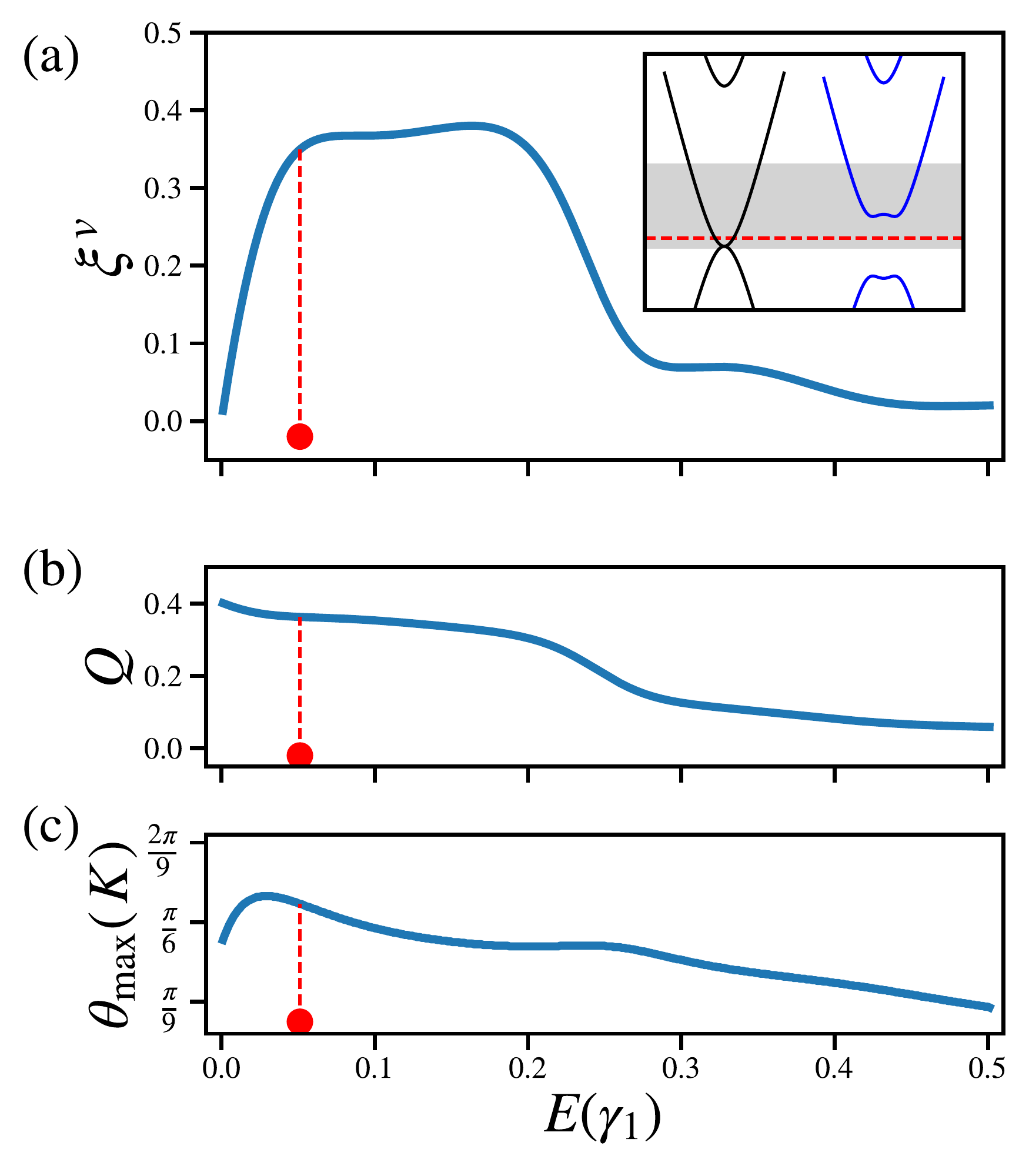}
	\caption{(a) Valley-scattering efficiency $\xi^v$, (b) total scattering efficiency $Q$ and (c) angle of maximum $K$ valley scattering $\theta_\mathrm{max} (K)$ for a biased dot with $R=5.0 \, l_0$ and $V_1=V_2=-0.2\,\gamma_1$. The red symbol and dashed lines show the energy discussed in the text and Fig. \ref{figredenergy}. The inset in panel (a) shows the alignment of bands outside (black) and inside (blue) the dot region, and the shaded area shows the energy range considered in the main panels. }
	\label{fig_edep}
\end{figure}

We begin by considering the energy and valley dependence of the far-field scattering from a biased dot.
For simplicity, all energies are given in units of $\gamma_1 = 0.38$ eV and lengths in units of $l_0 = \tfrac{h v_F}{\gamma_1} \approx 2 $ nm.
Fig. \ref{fig_edep}(a)--(b) show the valley ($\xi^v$) and total ($Q$) scattering efficiencies  for a dot with radius $R=5.0\, l_0$, and $\Delta = -0.4\, \gamma_1$, corresponding to layer potentials $V_1=-V_2=-0.2\,\gamma_1$. 
The inset in Fig. \ref{fig_edep}(a) shows the alignment of the band structures outside (left, black) and inside (right, blue) the dot, with the gray shaded area showing the energy range considered in the main panels.
Similar to the case of mass dots embedded in monolayer graphene\cite{Aktor2021}, the valley scattering maintains a consistent sign throughout the energy range.
This indicates that $K$ valley electrons are preferentially scattered in the $+y$ direction, and $K^\prime$ electrons in the $-y$ direction, within this energy range.
The valley scattering is most effective within the gap of the biased region, $|E| \lesssim 0.2 \gamma_1$, where both $\xi^v$ and $Q$, are reasonably constant before beginning to decay towards the conduction band edge. 
In Fig. \ref{fig_edep} (c), we plot $\theta_{\mathrm{max}}(K)$, the angle for which $j^K$ is maximum, \emph{i.e.} the preferred scattering angle for electrons from the $K$ valley.
We note that reasonably uniform deflection angle of $\approx 30^\circ$ throughout the gap region, with a reduction of $\theta_{\mathrm{max}}(K)$, indicating more forward scattering, at higher energies.
This is in contrast to the behaviour of mass dots in monolayer graphene, where the preferred scattering angle changes more dramatically, from being almost perpendicular to the incoming plane wave at low energies to much smaller angles near the band edge.\cite{Aktor2021}

\begin{figure}
\includegraphics[width=0.48\textwidth]{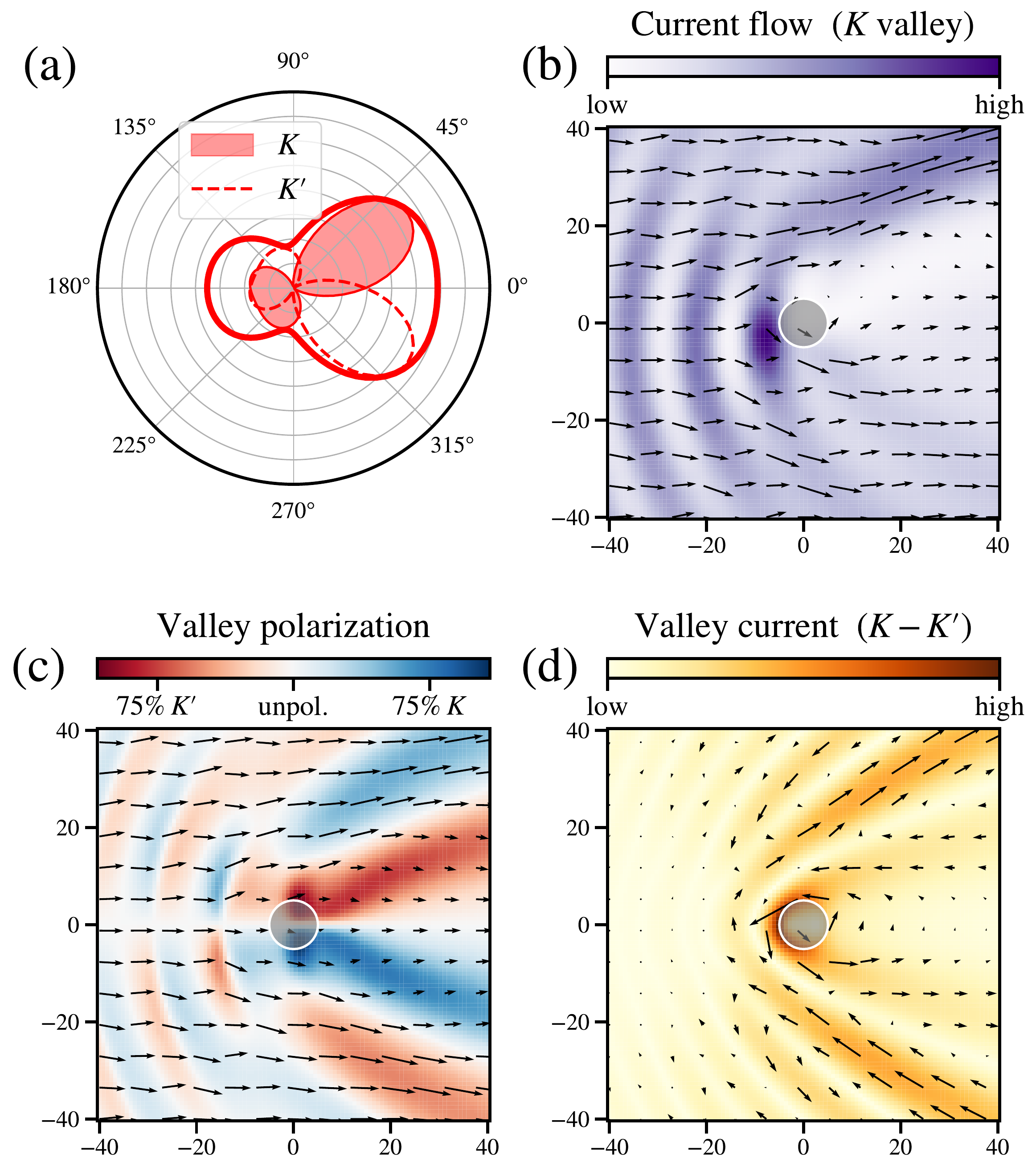}
	\caption{Valley splitting at a biased dot, with $R=5.0 \,l_0$, $V_1=-V_2=0.2\,\gamma_1$, at the energy $E=0.051\,\gamma_1$ shown by the red dot in Fig. \ref{fig_edep}. (a) Far-field angular dependence of scattering for each valley and the total current. (b) Local flow of $K$ valley electrons near the dot. (c) Total current flow (arrows) and valley polarization (colour) near the dot. (d) Valley current flow near the dot. }
	\label{figredenergy}
\end{figure}

We now consider in more detail how the valley and total electronic current are affected by a biased dot.
Fig. \ref{figredenergy} examines the far-field and local current behaviours for incoming electrons at an energy within the band gap of the dot.
The exact energy chosen ($E=0.051\gamma_1$) is denoted by the red dot in Fig. \ref{fig_edep}.
The polar plot in Fig. \ref{figredenergy}(a) shows the angular dependence of the far-field scattered current for the $K$ (shaded) and $K^\prime$ (dashed) valleys independently, and also the combined total current (bold curve).
The total current is primarily scattered in the forward direction, in a uniform beam between $-30^\circ \lesssim \theta \lesssim  30^\circ $.
However, we note that the individual valley contributions to this beam are anti-symmetric around the $x$-axis.
The two valleys contribute equally at $\theta=0$, so that current flow directly behind the dot is unpolarized.
However, the edges of the beam at $\theta \approx  \pm 30^\circ $ are strongly polarised, with the top (bottom) of the scattered beam consisting almost entirely of $K$ ($K^\prime$) electrons.

The preferred direction of $\theta \approx 30^\circ$ for $K$ electrons is in agreement with the $\theta_{\mathrm{max}}(K)$ calculation in Fig. \ref{fig_edep}(c), and is also evident in the current map for the $K$ valley in the vicinity of the dot, shown in Fig. \ref{figredenergy}(b).
The arrow length and background shading show the magnitude of the current, which for this valley is preferentially deflected above the dot, leading to the strong forward current visible in the upper-right part of the panel.
In addition, a periodic fluctuation in the magnitude of the current is noted in front of the dot, which can be associated with interference between incoming and back-scattered currents.
Both these local trends near the dot are consistent with the far-field angular behaviour shown in Fig. \ref{figredenergy}(a).

The behaviour of the $K^\prime$ current in both the far-field limit (dashed line in Fig. \ref{figredenergy}(a)) and the vicinity of the dot (not shown) is identical to that of the $K$ valley, but mirrored through the $x$-axis.
If an incoming $K$ valley current scatters in the $+y$ direction, or flows mostly in a clockwise direction around the dot as shown in Fig. \ref{figredenergy}(b), then the corresponding $K^\prime$ current scatters in the $-y$ direction and has a anti-clockwise flow pattern.
This symmetry is associated with the mode symmetry noted in Fig. \ref{figmodes}: each $K$ valley mode has a corresponding $K^\prime$ mode with equal magnitude but opposite angular momentum.
The total electronic current flow, being a sum of $K$ and $K^\prime$ contributions, is therefore symmetric around $y=0$.
This is true for both the far-field case, as shown by the bold red curve in Fig. \ref{figredenergy}(a), and by the arrows showing the total current flow around the dot in Fig. \ref{figredenergy}(c).
The color scale under the arrows in this plot shows the valley polarization of the current.
As expected, we see a region of blue shading in the upper-right part of the panel, corresponding to the the $K$ valley flow above the dot which leads to the far-field behaviour discussed earlier.
A corresponding region with red shading, and $K^\prime$ polarization, is seen in the bottom right of the panel.
However, regions with opposite and stronger polarization are noted at smaller angles behind the dot. 
In Fig. \ref{figredenergy}(c), a strong $K^\prime$ polarization can be seen at small positive angles beneath the prominent $K$ valley flow, and which coincides with a shadow in the $K$ valley behind the dot in Fig. \ref{figredenergy}(b).
Although the current in this region is strongly $K^\prime$-polarized, the magnitude of the current is small.
Furthermore, we noted that $K^\prime$ electrons in this region soon mix with $K$ electrons flowing under the dot, so that the valley polarization at small angles decays quickly as we move away from the dot.
This trend is also visible in the valley current, \emph{i.e.} the difference of the $K$ and $K^\prime$ contributions, near the dot shown Fig. \ref{figredenergy}(d).
A strong valley current is observed right at the dot edge, but its magnitude quickly decays behind the dot.
The only prominent non-vanishing flow enters from the bottom right, flows around the dot at a small separation, and exits again to the top right.
Again this coincides with the expected far-field behaviour.
The induced valley current is somewhat similar to that predicted for mass dots in monolayer graphene, except that in the monolayer case the valley current is more perpendicular to the incident wave.
Nonetheless, the bilayer result has a significant transverse component, and therefore scattering from multiple such dots will also give rise to an extrinsic valley Hall effect.
As it is driven by scattering of electrons at the Fermi surface, and not Berry curvature effects within the Fermi sea, this type of valley Hall effect can be directly connected to device measurements within non-local or similar geometries.\cite{Aktor2021}

\subsection{Size and strength of biased dot} \label{sec_size_strength}

\begin{figure}
\includegraphics[width=0.48\textwidth]{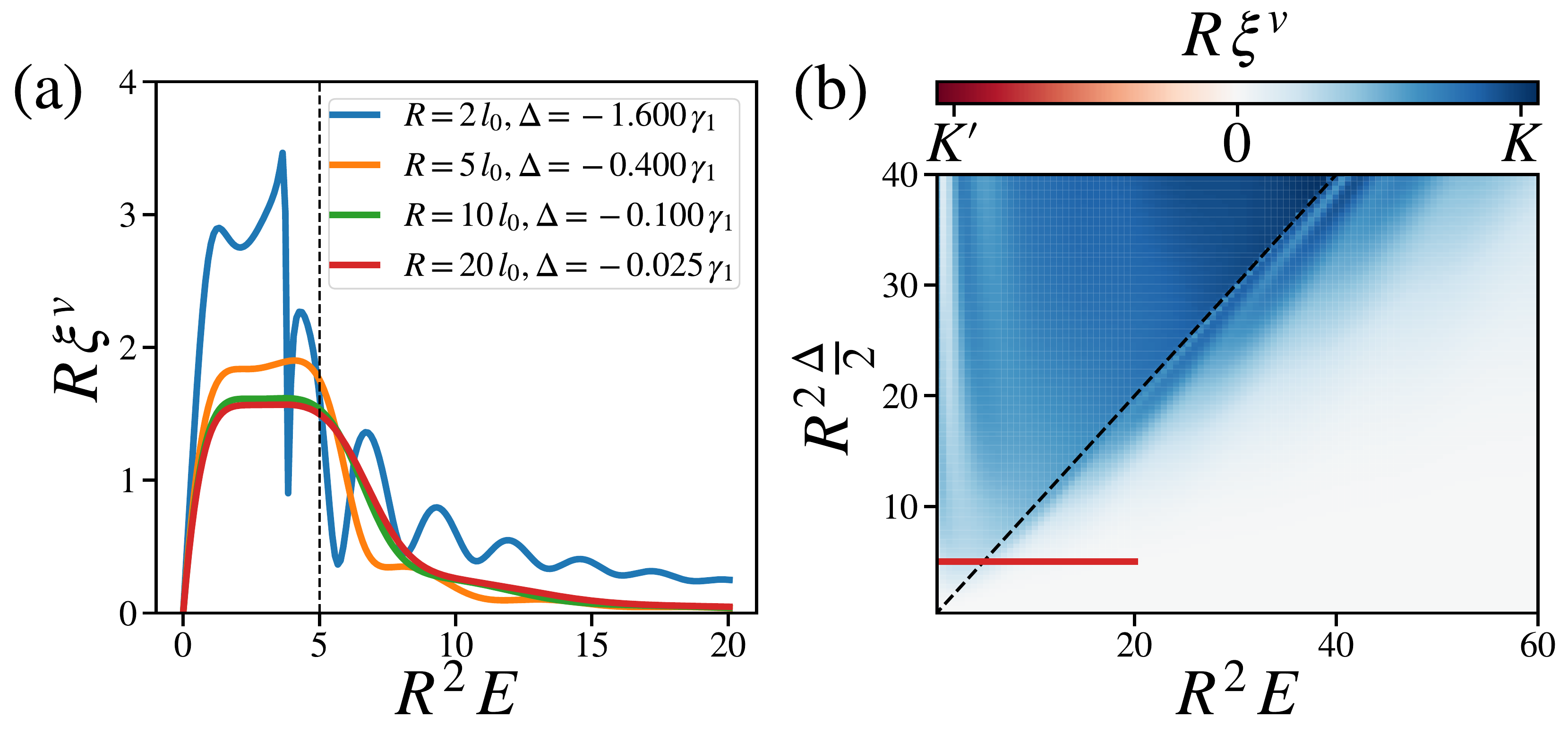}
	\caption{Scaling of valley polarization with interlayer potential strength and size of dot. (a) For $E, \Delta \ll \gamma_1$, $\xi^v$ tends towards a master curve for systems with the same value of $R^2 E$ and $R^2 \Delta$. (b) Master diagram for valley scattering for small energies and interlayer potentials. Note that the valley scattering has a consistent sign (i.e. color) within the dot gap. The red curve in (a) can be viewed as a slice of this plot, with its location shown by the solid red line. The dashed lines in both panels show the expected band edge at $E=\frac{\Delta}{2}$. }
	\label{figscaling}
\end{figure}

Having explored the energy, valley and angular dependence of scattering from a particular dot, we now consider how this depends on the dot size and the magnitude of the potential.
We note that, similar to the case of mass dots in monolayer graphene, the energy and size dependencies are related.
This is because the majority of quantities depend on these only through terms like $k R$, which can be kept constant for dots with different radii by adjusting the wavevector $k$ accordingly.
For monolayer systems, this can be achieved by scaling all energy terms ($E$, $\Delta$) inversely to the change in length scale, as $E \sim k$.
In bilayer graphene systems, the scaling is instead $E \sim k^2$, which suggests that we can still produce a ``master diagram'' showing the valley-scattering behaviour over a range of sizes and energy scales.
However, in the bilayer case, it is not straightforward to simply scale up or down all the energy terms.
The interlayer hopping parameter $\gamma_1$ is fixed and, unlike all the other energy terms in monolayer and bilayer graphene, is not tunable using local or global gates.
The dispersion relations in Eq. \eqref{eq:V1V2dispersion} and \eqref{eqE0} do not give the same physics simply by rescaling the dot sizes and potentials, as in the case of monolayer dots.
This is shown clearly in Fig. \ref{figscaling}(a), where we plot the appropriately scaled valley scattering efficiency for four different systems where the interlayer potentials have been scaled $\Delta \sim \tfrac{1}{R^2}$. 
The orange curve here corresponds to the result in Fig. \ref{fig_edep}(a), and we note the different behaviour seen for smaller dots with stronger potentials (blue curve) and larger dots with weaker potentials (green and red curves).
It is clear that the valley scattering in bilayer systems is pinned to the exact values of $E$ and $\Delta$ considered due to the fixed value of $\gamma_1$.
This is particularly true for larger values of $\Delta$, such as the blue curve in Fig. \ref{figscaling}(a).
As the magnitudes of $V_1$ and $V_2$ become comparable to $\gamma_1$, the effect of the interlayer potential changes.
The shapes of the bands change, leading to the characteristic ``Mexican Hat'' shape and a shift of the band edge away from the $K$ points. \cite{McCann2013} 
Furthermore, the band gap in dual-gated BLG saturates at $2\gamma_1$, instead of increasing continuously with the interlayer potential.
This is different to a sublattice-dependent mass term in SLG, where the band gap continues to increase with the mass.

For small values of the interlayer potential, $|\Delta| \ll \gamma_1$, the band gap behaves similarly to one induced by a mass term with a band edge at $E\approx \frac{\Delta}{2}$.
In this limit, we do find uniform behaviour if the dot size and potentials are scaled accordingly, as the wavevector depends only very weakly on $\gamma_1$.
This is shown clearly by the green ($\Delta=-0.1\gamma_1$) and red ($\Delta=-0.025\gamma_1$) curves in Fig. \ref{figscaling}(a), which coincide almost exactly when the axes are scaled.
It is therefore possible to produce a master diagram of valley scattering effects for systems with small interlayer potentials, as shown in Fig. \ref{figscaling}(b).
The plot is calculated for an $R=20 l_0$ dot over a wide range of energy and interlayer potentials.
The range considered by red curve in Fig. \ref{figscaling}(a) is shown by the solid red line.
The approximate band edge ($E=\tfrac{\Delta}{2}$) is shown by the dashed black line, and we note that strong valley-scattering is seen for $E<\tfrac{\Delta}{2}$, with the valley scattering efficiency quickly decaying towards zero for larger energies.
We also note that the sign (colour) of the valley scattering is uniformly positive (blue), corresponding to a preferential scattering of $K$ electrons in the $+y$ direction, over the energies considered.
Therefore, scattering from dual-gated dots in bilayer graphene appears to be a robust mechanism for valley current generation: the qualitative effect is similar over a range of energies and interlayer potentials.
While mass dots in monolayer graphene might induce a stronger effect, it is far more feasible experimentally to control an asymmetry between two layers than between sites from two sublattices.
Furthermore, the angular dependence is more uniform in biased dots, which is of vital importance for devices relying on consecutive scattering from multiple dots to amplify the valley scattering effects.

\subsection{Asymmetric potentials} \label{sec_asymm}

\begin{figure}
\includegraphics[width=0.48\textwidth]{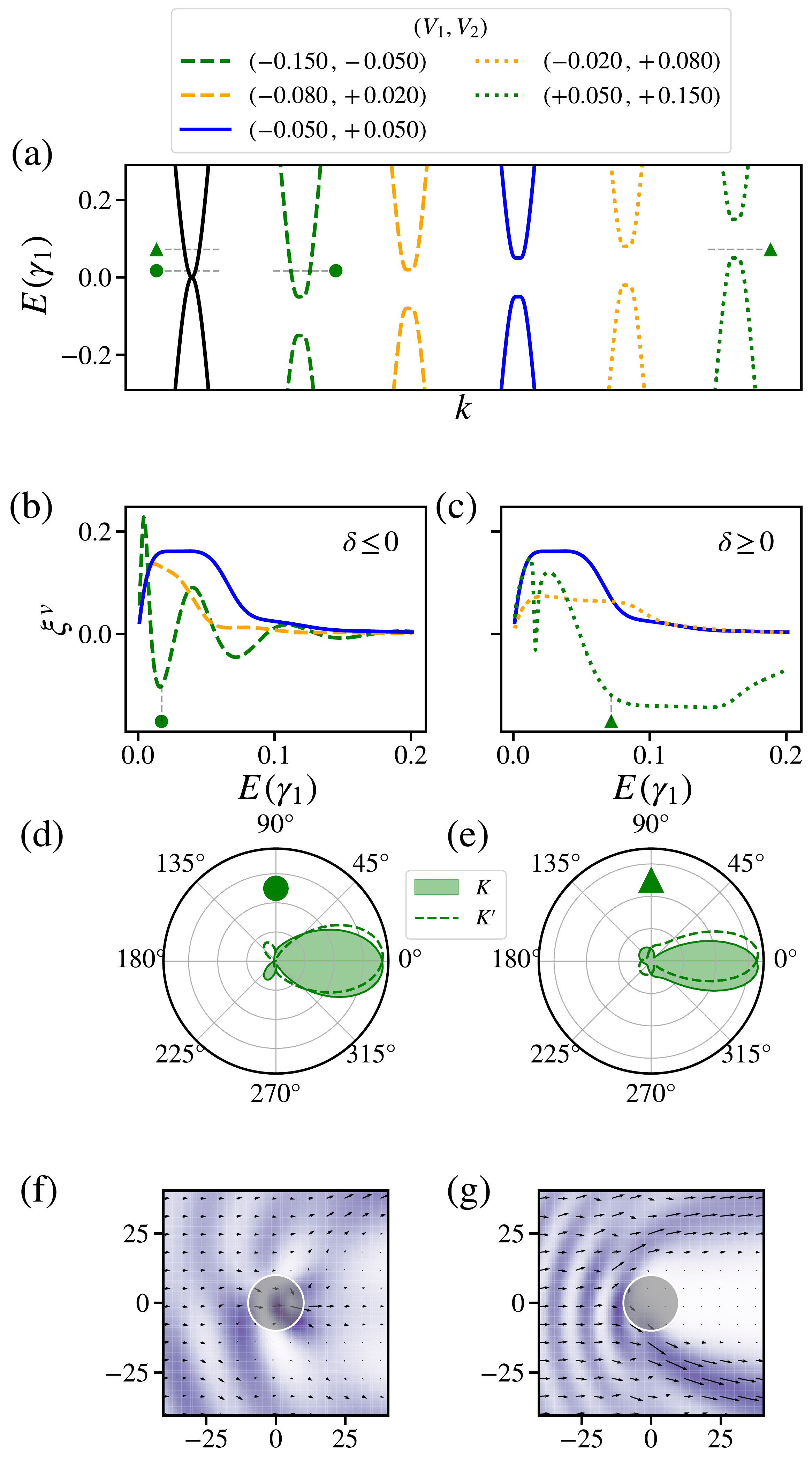}
	\caption{Scattering from $R=10 \, l_0$ dots with the same interlayer potential $\Delta$, but a different band centre shift $\delta$. 
	(a) Alignment of the band structure outside the dot (black) with the various dots considered. The solid blue curve corresponds to the $\delta=0$ case.
	(b),(c) show the valley scattering efficiency for dots where the band centre is shifted down or up, respectively.
	(d),(e) show the far-field scattering, and (f),(g) the local $K$ valley current flow, for specific cases shown by the corresponding symbols in the upper panels. }
	\label{figV1V2}
\end{figure}

A dual-gated setup also allows us to shift the band centre $\epsilon_c$ of the dot region relative to the surrounding sheet. 
In this case, the potentials on the two layers are no longer of equal magnitude, so that $V_1 \ne - V_2$ and $\delta \ne 0$. 
We consider an  $R=10 l_0$ dot with $\Delta=-0.1\gamma_1$, similar to that represented by the green curve in Fig. \ref{figscaling}(a) for $\delta=0$.
Fig. \ref{figV1V2}(a) shows how applying a shift affects the alignment of bands outside the dot (black) and inside the dot, with the solid blue curve corresponding to the $\delta=0$ case.
Both downward (dashed) and upward (dotted) shifts of the band centre relative to the surrounding sheet are considered.
One immediate consequence is that incoming low-energy electrons no longer necessarily coincide with the gap of the dot, but can interact with states in the conduction (valence) band for a large enough negative (positive) value of $\delta$.
This occurs, for example, for the energy marked by the circle symbol, which is inside the conduction band for a dot with $\delta = -0.1 \gamma_1$ (green, dashed bands).

In Fig. \ref{figV1V2}(b),(c), we examine the effect of negative and positive shifts, respectively, on the valley scattering efficiency. 
In both cases the solid blue curve corresponds to the unshifted case, with the orange (green) curve corresponding to $|\delta| = 0.03\gamma_1$ ($0.1\gamma_1$). 
For small shifts, the qualitative behaviour is largely unchanged.
The orange, dashed curve in  Fig. \ref{figV1V2}(b) shows that a dot with a small negative shift gives similar valley splitting to the unshifted case at low energies, but that it decays faster as energy is increased.
This is associated with the earlier onset of the conduction band edge, which also led to the decay of $\xi^v$ in the $\delta =0$ case.
A similar effect is seen for a small upward shift (orange, dotted curves in Fig. \ref{figV1V2}(c)), where the conduction band onset and valley scattering decay are now shifted upwards in energy.  
The decrease in the magnitude of $\xi^v$ for $\delta=+0.03$ is associated with an increase in a backscattered component with opposite polarisation.
A similar, but smaller, effect can be seen in the far-field behaviour for the unshifted dot in Fig. \ref{figredenergy}(a), where we both a strong forward-scattered $K$ lobe and a smaller back-scattered $K^\prime$ lobe are present in the upper half-plane. 
The behaviour of the forward-scattered $K$ component for the $\delta=+0.03$ dot is qualitatively similar to the unshifted case, but a relative increase in the backscattered $K^\prime$ component leads to an overall decrease in $\xi^v$.
While the $\delta=+0.03$ dot is a weaker generator of transverse valley currents than the unshifted dot, as a valley polarizer of left-to-right current it will perform similarly.

Larger shifts of the band centre in the dot, shown by the green curves in Fig. \ref{figV1V2}(a)--(c), show qualitatively different behaviour.
For $\delta< - \tfrac{\Delta}{2}$ (dashed green curve), there is no overlap between the low energy conduction band outside the dot and the gap inside the dot.
The corresponding plateau in valley scattering efficiency is also absent, and has been replaced by a highly oscillatory curve in Fig. \ref{figV1V2}(b).
While this might suggest highly tunable valley behaviour, the majority of the forward-scattered has very little deflection, as shown by the far-field  plot in  Fig. \ref{figV1V2}(d).
The local $K$ valley map for this case in Fig. \ref{figV1V2}(f) shows that a significant part of the current is guided though the dot, which is not surprising as there are states available in the dot at this energy.
This leads to a smaller angular deflection, and only minor splitting of the valleys.
The valley efficiency for a large positive shift (dotted green curve in Fig. \ref{figV1V2}(c)) shows an inversion of the valley polarization compared to the unshifted case.
However, at energies in this range (such as that denoted by the triangle) the far-field and local behaviour in Fig. \ref{figV1V2}(e),(g) show that there is not a substantial difference between the two valleys.
While the current flows around the dot, the flow for both valleys is largely symmetric around the $y$-axis, with only slightly more $K$ electrons flowing under the dot than over it.

Overall the valley-splitting effect tends to be weakened by the addition of an energy shift.
However, the same qualitative valley-splitting behaviour that was present for unshifted dots also emerges in the presnce of a small shift.
This suggests that the desired effect should be robust in an experimental setup as it is not critically dependent on exactly equal and opposite potential on the two layers.
While a large shift in energy can change the sign of the induced valley current, it becomes more difficult to distinguish the individual $K$ and $K^\prime$ valley beams.
As the shift is increased, the gap moves further from the low-energy range of interest, and the system resembles a valley-neutral potential dot, such as those considered in previous works.\cite{Peterfalvi2009}
A direct inversion of the layer potentials in the dot region, \emph{i.e.} $\Delta \rightarrow - \Delta$, is the most effective method to flip the sign of the valley current as it maintains the features discussed in Figs. \ref{fig_edep} and \ref{figredenergy} while swapping the roles of the $K$ and $K^\prime$ valleys.

\subsection{Bilayer dots embedded in single layer graphene} \label{sec_SLG}
Biased dots appear a robust platform for valley current generation in bilayer graphene. 
They replicate many key features of mass dots in SLG,\cite{Aktor2021} but in a setup that is easier to fabricate and tune in experiment.
With this in mind, we now consider if the dual-gated setup can also be employed to induce valleytronic effects in a single layer.
We consider a graphene sheet with a small bilayer region where dual-gating can be applied, so that our system consists of both a continuous and a finite layer.
In solving the scattering problem, we proceed as in Section \ref{secmethods}, but replacing the incoming and reflected wavefunctions with their SLG equivalents.
For example, the wavevector is now given by $k = |\tilde{E}|$ and the individual mode solutions in Eq. \eqref{eq:psimJdef} take the form   
\begin{align}
    \psi^{0,J}_m(kr,\theta) \sim \begin{pmatrix}
    -i J_{m-\tau}(kr) e^{i(m-\tau)\theta} \\
    \eta \tau J_m(kr) e^{im\theta} 
    \end{pmatrix}
\end{align}
outside the dot.
In applying boundary conditions at the dot edge, we follow the approach discussed in Ref. \onlinecite{linearBL} for linear bilayer barriers in a graphene sheet.
(Note that the site indexing convention is slightly different in this work.)
The bottom layer of the BLG region is continuous with the single layer outside, allowing us to write the first two conditions:
\begin{align}
    \psi^A_\mathrm{inc} (k, \mathbf{R}) + \psi^A_\mathrm{sc} (k, \mathbf{R}) = \psi^{a_1}_\mathrm{tr} (k, \mathbf{R})\\
    \psi^B_\mathrm{inc} (k, \mathbf{R}) + \psi^B_\mathrm{sc} (k, \mathbf{R}) = \psi^{B_1}_\mathrm{tr} (k, \mathbf{R}) \,,
\end{align}
where $A$ and $B$ refer to the sublattices in SLG.
The condition applied to the top layer determines the edge geometry of the finite dot.
A realistic circular dot would have a combination of armchair and zigzag edges of different types,\cite{Bhowmick2008, Heiskanen2008, Wimmer2010, Power2014} but it is not feasible to consider this within the continuum approach being used here.
We instead consider two different zigzag edge types, `ZZb2' and `ZZA2', shown in Fig. \ref{setup}(d), which are applied uniformly around the circumference of the dot.
Qualitative similarities between these cases should indicate universal behaviour, whereas differences between them will establish the relative importance of edge effects in finite dots.
The zigzag edge types are imposed by setting the wavefunction on one of the two sublattices in the finite layer to zero at the dot's edge
\begin{align}
\psi^{A_2}_\mathrm{tr} (k, \mathbf{R}) & =0 \quad\quad \mathrm{(ZZb2 \; edge)} \\
 \psi^{b_2}_\mathrm{tr} (k, \mathbf{R})& =0 \quad\quad \mathrm{(ZZA2 \; edge)}
\end{align}
so that the edge of the finite layers consists solely of atoms from the other sublattice.
We denote the edge type by this remaining sublattice, so that `ZZb2' refers to a (zigzag) edge of non-dimer sites from the B sublattice, whereas `ZZA2' is composed of dimer A atoms. 
We can now calculate the scattering coefficients and efficiencies using an approach analogous to Eq. \eqref{eqQ} - \eqref{eqxi}.

\begin{figure}
\includegraphics[width=0.48\textwidth]{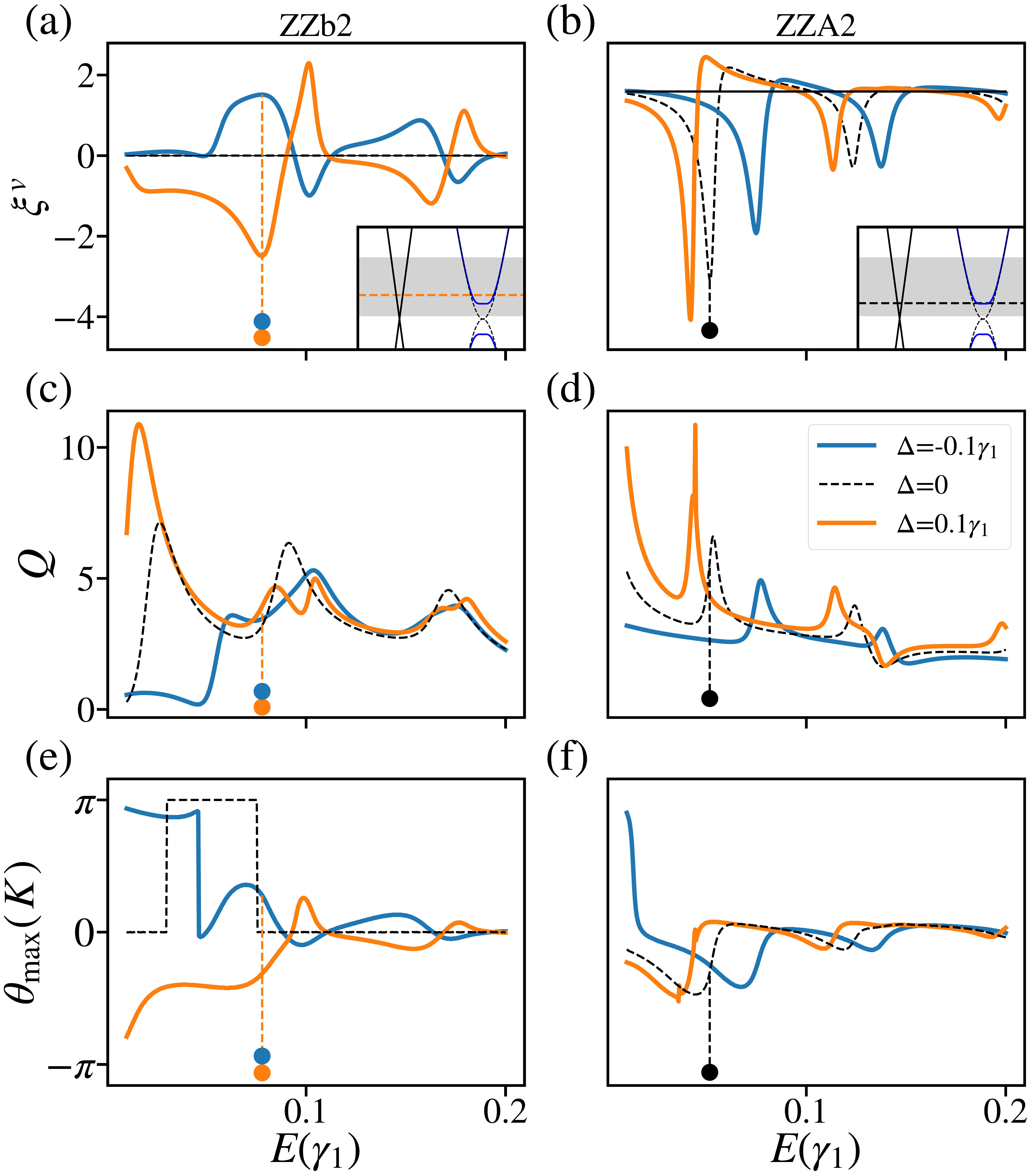}
	\caption{	Scattering from bilayer dots in SLG with ZZb2 (left) and ZZA1 (right) edges for a negative (blue), zero (black, dashed) or positive (orange) interlayer potential applied in the dot region. (a),(b) show valley-scattering efficiency $\xi^v$. (c),(d) shows total scattering efficiency $Q$, and (e),(f) the angle of maximum $K$ valley scattering $\theta_\mathrm{max} (K)$. The  dots highlight particular energies discussed in the text and Fig. \ref{individual_results}. The insets in (a) and (b) shows the alignment of bands outside (black, left) and inside the bilayer dot, with (blue, right) and without mass (dashed, right). 
	}
	\label{mlbl_results}
\end{figure}

The two layers of the system are no longer equivalent, so inverting the interlayer potential can have effects beyond swapping the behaviour of the two valleys. 
In Fig \ref{mlbl_results}, we examine the behaviour of an $R=10 l_0$ dot at small energies not only for negative (blue) interlayer potentials  but also for the equivalent positive (orange) case.
Unlike the bilayer case, the finite dot also acts as a scatterer here even in the absence of any potential terms.
The $\Delta=0$ case, shown by the black dashed line, allows us to determine the scattering effects introduced by the presence of the finite bilayer region separately to those of the mass term introduced by the interlayer potential.
We first note that the behaviour in Fig \ref{mlbl_results} is in general much less uniform than the BLG case shown in Fig. \ref{fig_edep}.

\begin{figure}
\includegraphics[width=0.48\textwidth]{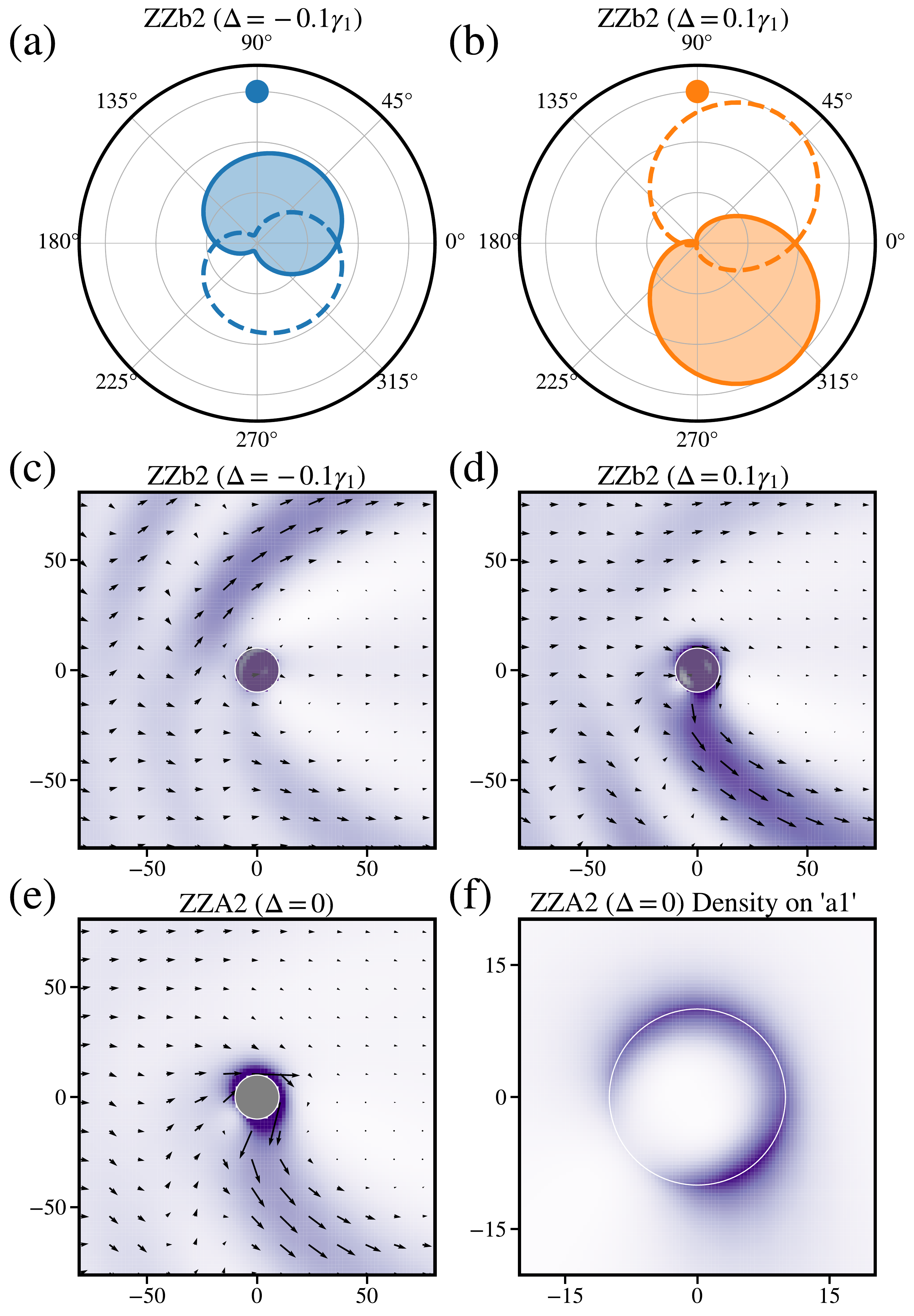}
 	\caption{(a) Far-field scattering for a dot with a ZZb2 boundary and $\Delta=-0.1 \gamma_1$ for $E=0.078\gamma_1$ (blue curve and dot in \ref{mlbl_results}(a)). (b) Corresponding far-field scattering for $\Delta=0.1 \gamma_1$ (orange curve and dot in Fig. \ref{mlbl_results}(a)). 
 	 (c), (d) Corresponding local current flows for the $K$ valley for the systems and energies in panels (a) and (b).
 	 (e) Local $K$ valley flow for a ZZA2 boundary dot with $\Delta=0$ for $E=0.052\gamma_1$ (black dot and line in Fig. \ref{mlbl_results}(b)).
 	 The very high local current inside the dot is hidden to aid visualization. 
 	 (f) Electron density on 'a1' sites, \emph{i.e.} the non-dimer sublattice sites on the continuous layer, for the system in panel (e).
 	}
 	\label{individual_results}
\end{figure}

For the ZZb2 boundary, we see from Fig. \ref{mlbl_results} (a),(e) that the valley scattering efficiencies and preferred scattering angles have almost equal and opposite behaviour for positive and negative masses.
This is particularly true for higher energies, where the overall scattering efficiencies $Q$ for both cases in Fig. \ref{mlbl_results} (c) also coincide, mimicking the expected behaviour for biased dots in a bilayer.
We note from Fig. \ref{mlbl_results}(e) that at energies such as that denoted by the circles, a large angular separation of valleys is achieved for both cases.
This is seen more clearly in the individual far-field angular plots in Fig. \ref{individual_results}(a),(b), which show large transverse scattering more akin to mass dots in SLG than to biased dots in BLG. 
The $\Delta=0$ dot has no far-field valley polarisation, but does show a similar overall scattering efficiency $Q$ as the biased dots.  

The results for the corresponding dots with ZZA2 edges in the right hand panels of Fig. \ref{mlbl_results} are strikingly different. 
All three dots show extremely similar behaviour, but with small relative energy shifts.
The equal and opposite behaviour expected when the interlayer potential is inverted is absent, and furthermore, a significant valley scattering occurs even in the absence of a mass term.
We note that although the symmetry between positive and negative mass dots is broken, each individual dot produces symmetric effects for the $K$ and $K^\prime$ valleys.
The stark contrast between the ZZb2 and ZZA2-edged dots can be understood in terms of the localised states that arise in these systems.\cite{Castro2008, linearBL, Mirzakhani2018}

The ZZb2 edge gives rise to an edge state which resides solely on b2 sites, \emph{i.e.} on the non-dimer edge sites of the finite layer. \cite{linearBL}
This state has little effect on the overall scattering characteristics of the dot as it resides on sites that do not directly couple to the surrounding graphene layer.
Instead, we see that the asymmetry between positive and negative interlayer potentials arises due to confined states.
These have been widely studied in a range of monolayer and bilayer quantum dots,\cite{Heiskanen2008, Pereira2007, Bardarson2009, Xavier2010, Guclu2011, Sebrowski2013, DaCosta2014, Mirzakhani2016, Gutierrez2016} and can have finite weight on both layers and sublattices.
Examining the local current flow for different masses reveals that although the behaviour is approximately equal and opposite in the far-field, this is not necessarily the case near the dot.
In the $\Delta=0.1 \gamma_1$ case the current is deflected as it flows through the dot (Fig. \ref{individual_results}(c)) whereas in the $\Delta=-0.1 \gamma_1$ case the strong valley current is scattered further from the dot Fig. \ref{individual_results}(d)).
Unlike in a pure bilayer system, the interaction with confined states here is different for positive and negative masses due to the inequivalency of the two layers.

The situation is very different for a ZZA2 edge, where the associated localised edge state resides on dimer A2 sites in the finite layer and on non-dimer a1 sites in the continuous layer.
The finite weight of this state on the continuous layer, shown explicitly in Fig. \ref{individual_results}(f), leads to a much stronger coupling with incoming waves.
This gives rise to a strong valley-splitting even in the absence of an interlayer potential, as shown by the black dashed line in Fig. \ref{mlbl_results}(b).
We see from Fig. \ref{individual_results}(e) that this state `guides' $K$ electrons around the edge of the dot, giving rise to a strong transverse scattering.
The net effect is similar to that with an interlayer potential in Fig. \ref{individual_results}(d).
The edge state effect is so dominant in the ZZA2 system that the addition of an interlayer potential of either sign does not change the overall direction of the valley splitting.
The same qualitative far-field and local current behaviour is seen for all three curves.
The main effect of the interlayer potential is to shift the peak position slightly in energy, with the shift direction given by the sign of the potential on the continuous layer. 

Our results suggest that although strong valley splitting can occur at bilayer dots embedded in a graphene sheet, the behaviour is less predictable than for mass dots in SLG or dual-gated dots in BLG.
The cases investigated above show that interlayer potentials and edge states can both induce valley splitting, but that the dominant contribution depends on the edge geometry of the finite layer.
We note that realistic dots will likely combine a mixture of different edge types.
Circular dots, for example, contain a mix of zigzag and armchair edge segments. \cite{Bhowmick2008}
While a hexagonal dot could contain only zigzag edges,\cite{Guclu2011} these would be a mixture of ZZA2 and ZZb2 edge types,\cite{linearBL} and a perfectly triangular dot would be required to restrict edge atoms on the finite layer to a single sublattice.\cite{Guclu2011} 
Therefore, for most BLG dots embedded in SLG the flow of valley current will be dictated by an interplay between edge- and gate-induced valley splitting.

\section{Conclusion}
We have demonstrated the emergence of a robust valley-splitting mechanism, and resulting valley current generation, at biased bilayer graphene dots due to the presence of an interlayer potential.
These effects occur whether the biased dot is embedded in a bilayer or a single layer graphene sheet, but with important qualitative differences.
In a bilayer sheet, the valley-dependent scattering is quite robust as the energy is varied, similar to the case of sublattice-dependent mass dots embedded in a single layer.\cite{Aktor2021}
This system is therefore an ideal platform for the experimental study of valley Hall effects, offering not only robust valley current generation but also an extrinsic, Fermi surface mechanism allowing direct connection with quantum transport simulations and the possibility to tune the interlayer potential strength.
Valley-dependent scattering also arises if a biased bilayer dot is placed in a single layer of graphene, but the effect depends strongly on the exact electron energy and dot edge geometry.
The current flow in this case is an interplay between the edge and confined states in the dot and the interlayer potential applied.

Graphene systems with non-uniform mass distributions are an extremely promising route towards the generation and manipulation of valley currents.
Bilayer graphene offers the distinct advantage of an external knob to vary the strength of the mass term, which is very difficult to achieve in single layer systems.
The use of dual-gated setups\cite{dotsKurzmann2019, dotsKurzmann2019a, dotsBanszerus2020, dotsBanszerus2020a, dotsGe2020, dotsGarreis2021, Overweg2018, Kraft2018, Lee2020} and gate patterning\cite{huber2020tunable} allows the fabrication of devices with multiple dots, so that consecutive scattering from dots can amplify valley polarization effects or induce measurable non-local resistance signals.
Finally, introducing a finite twist between the two layers is a natural extension of this work which offers further degrees of tunability, both in the spatial fluctuation of mass terms due to different stackings and the different confined or edge states in the dot regions.\cite{Mirzakhani2020, Bucko2021}

\begin{acknowledgments}
The authors wish to acknowledge the support of the Irish Centre for High-End Computing (ICHEC).
S.R.P. acknowledges funding from the Irish Research Council under the Laureate awards programme.
\end{acknowledgments}

\end{document}